\documentclass[useAMS,usenatbib]{mn2e}

\usepackage{graphicx}
\usepackage{footmisc}
\usepackage{color} 
\usepackage{rotating}
\usepackage{amssymb}
\usepackage{multirow} 
\usepackage{booktabs} 

\newcommand{\lum}{{\rm erg\,s^{-1}}}
\newcommand{\flux}{{\rm erg\,s^{-1}cm^{-2}}}
\newcommand{\kms}{{\rm km\,s^{-1}}}
\newcommand{\nh}{{\rm cm^{-2}}}
\newcommand{\red}{$z$}

\newcommand{\aap}{A\&A}
\newcommand{\aj}{AJ}
\newcommand{\apj}{ApJ}
\newcommand{\apjs}{ApJS}

\newcommand{\mnras}{MNRAS}
\newcommand{\araa}{ARA\&A}

\newcommand{\xmm}{\textit{XMM-Newton }}
\newcommand{\chandra}{\textit{Chandra }}

\newcommand{\sgm}{$\sigma$ }                      %
\newcommand{\asec}{$\arcsec$}                      %
\newcommand{\mic}{$\mu$m }                      %
\newcommand{\nmic}{$\mu$m}                      %
\newcommand{\wise}{$\rm WISE$ }                      %
\newcommand{\sn}{SNR}                      %
\newcommand{\buxs}{\texttt{BUXS} }

\title[AGN selection with WISE]{Using the Bright Ultra-Hard \xmm Survey to define an IR selection of luminous AGN based on \wise colours} 
\author[Mateos et al.] {S. Mateos$^{1,2}$\thanks{E-mail:
    mateos@ifca.unican.es}, A. Alonso-Herrero$^{1}$\thanks{Augusto G. Linares Senior Research Fellow}, F. J. Carrera$^{1}$, A. Blain$^{2}$, M. G. Watson$^{2}$, 
\newauthor 
X. Barcons$^{1}$, V. Braito$^{3}$, P. Severgnini$^{4}$, J. L. Donley$^{5}$ and D. Stern$^{6}$  
\smallskip \\
\footnotesize
$^{1}$ Instituto de F\'isica de Cantabria (CSIC-Universidad de Cantabria), 39005, Santander, Spain\\
$^{2}$ Physics and Astronomy, University of Leicester, University Road, Leicester LE1 7RH, UK\\
$^{3}$ Istituto Nazionale di Astrofisica - Osservatorio Astronomico di Brera, Via Bianchi 46 I-23807 Merate (LC), Italy \\
$^{4}$ INAF-Osservatorio Astronomico di Brera, via Brera 28, 20121 Milano, Italy  \\
$^{5}$ Los Alamos National Laboratory, Los Alamos, NM 87545, USA \\
$^{6}$ Jet Propulsion Laboratory, California Institute of Technology, Pasadena, CA 91109, USA}

\begin{document}

\date{Accepted 2012 July 31. Received 2012 July 27; in original form 2012 June 26}

\pagerange{\pageref{firstpage}--\pageref{lastpage}} \pubyear{2012}

\maketitle

\label{firstpage}

\begin{abstract}
We present a highly complete and reliable mid-infrared (MIR) colour
selection of luminous AGN candidates using the 3.4, 4.6, and 12
\mic bands of the \wise survey.  The MIR colour wedge was defined
using the wide-angle Bright Ultra-Hard \xmm Survey ({\tt BUXS}), one
of the largest complete flux-limited samples of bright (${\tt {\it
    f}_{4.5-10\,keV} > 6\,x\,10^{-14} \flux}$) ``ultra-hard'' (4.5-10
keV) X-ray selected AGN to date.  {\tt BUXS} includes 258 objects
detected over a total sky area of 44.43 deg$^2$ of which 251 are
spectroscopically identified and classified, with 145 being type-1 AGN
and 106 type-2 AGN. Our technique is designed to select objects
with red MIR power-law spectral energy distributions (SED) in the
three shortest bands of \wise and properly accounts for the errors in
the photometry and deviations of the MIR SEDs from a pure
power-law. The completeness of the MIR selection is a strong function
of luminosity. At ${\tt {\it L}_{2-10\,keV}>10^{44}\lum}$, where the
AGN is expected to dominate the MIR emission, $97.1_{-4.8}^{+2.2}\%$
and $76.5_{-18.4}^{+13.3}\%$ of the \buxs type-1 and type-2 AGN meet
the selection. Our technique shows one of the highest reliability and
efficiency of detection of the X-ray selected luminous AGN population
with \wise amongst those in the literature. In the area covered by the
\buxs survey our selection identifies 2755 AGN candidates detected
with \sn$\geq$5 in the three shorter wavelength bands of {\rm WISE}
with 38.5\% having a detection at 2-10 keV X-ray energies. We also
analyzed the possibility of including the 22{$\mu$m} \wise band to
select AGN candidates, but neither the completeness nor the
reliability of the selection improves. This is likely due to both the
significantly shallower depth at 22\mic compared with the first three
bands of \wise and star-formation contributing to the 22\mic emission
at the \wise 22\mic sensitivity.
\end{abstract}


\section{Introduction}

There is strong observational evidence that active galactic nuclei
(AGN) play an important role in the formation and growth of galaxies
(e.g. \citealt{magorrian98}). Most supermassive black hole growth
takes place during an obscured quasar phase, as suggested by the
integrated energy density of the cosmic X-ray
background \citep{fabian99}. To understand the evolution of galaxies
and to trace the energy output due to accretion and its cosmological
evolution, it is critical to map the history of obscured accretion.

X-ray surveys with \xmm and {\textit{Chandra}} at energies $<$10 keV
are sensitive to all but the most heavily obscured AGN
(e.g. \citealt{ceca08}). In Compton-thick AGN (rest-frame column
densities exceeding ${\rm N_H\simeq 1.5\times10^{24}\,cm^{-2}}$) the
observed flux below 10 keV can be as low as a few \% of the intrinsic
nuclear flux. In the Compton-thick regime the high energy photons that
survive the photoelectric absorption get scattered in the absorber
losing part of their energy (Compton down-scattering). This is an
important effect that can significantly suppress the transmitted
continuum (\citealt{matt02}; \citealt{murphy09};
\citealt{yaqoob10}). The ongoing Swift/BAT and INTEGRAL/IBIS all-sky
surveys at energies 15-200 keV are providing the least biased samples
of absorbed AGN in the local Universe (e.g. \citealt{bird07};
\citealt{tueller08}; \citealt{winter09}; \citealt{burlon11}). However,
even these surveys are biased against the most heavily absorbed
Compton-thick AGN \citep{burlon11}.

Surveys at mid-infrared (hereafter MIR) wavelengths
($\gtrsim$5{$\mu$m}) are much less affected by extinction since the
obscuring dust re-emits the nuclear optical-to-X-ray radiation at
infrared wavelengths. Clumpy torus models predict nearly isotropic
emission in the MIR at wavelengths $\gtrsim$12{$\mu$m}
(\citealt{nenkova08}). Thus, MIR-based surveys (or the combination of
MIR and data at shorter wavelengths) can potentially trace the elusive
obscured accretion missed by hard X-ray surveys
(e.g. \citealt{daddi07}; \citealt{fiore08};
\citealt{georgantopoulos08}; \citealt{fiore09};
\citealt{severgnini12}). For example, it has been claimed that objects
showing excess emission at $\sim$24\mic over that expected from star
formation, termed "infrared-excess galaxies", might host heavily
obscured and Compton-thick AGN (e.g. \citealt{fiore08};
\citealt{fiore09}). However the exact contribution of heavily obscured
AGN to the infrared-excess galaxy population remains an open issue
(e.g. \citealt{alexander11}). Several MIR-based AGN selection
techniques have been developed with data from the Spitzer Space
Telescope Infrared Array Camera (IRAC; \citealt{fazio04}) using
colours and power-law selection (\citealt{lacy04}; \citealt{stern05};
\citealt{alonso06}; \citealt{donley08,donley12}). These techniques are
very effective and reliable.  Galaxies dominated by AGN emission
typically exhibit a characteristic red power-law spectral energy
distribution (SED) in the MIR ($f_\nu\propto\nu^{\alpha}$ with
$\alpha\leq$$-$0.5; \citealt{alonso06}).  Thus, MIR power-law selection
provides the cleanest samples of luminous AGN
(e.g. \citealt{donley08}). However, this technique is very sensitive
to the reliability of the estimated photometric errors
\citep{donley12}.

The Wide-field Infrared Survey Explorer ({\rm WISE}) has now completed
the first sensitive ($\sim$100-1000$\times$ deeper than {\rm IRAS})
coverage of the entire sky in the
MIR\footnote{https://ceres.ipac.caltech.edu/}\citep{wright10}. Several
colour-based regions, aimed at identifying luminous AGN, have already
been proposed. These works have shown that \wise can robustly separate
AGN from normal galaxies and stars (e.g. \citealt{assef10};
\citealt{jarrett11}; \citealt{stern12}). \wise will be extremely
efficient in identifying the rare highly luminous AGN up to the
crucial epoch when the accretion power of the Universe peaked
(\red$\sim$1-2). The all-sky \wise survey will complement the deep
Spitzer surveys, aimed to characterize the accretion phenomenon in the
distant Universe.

This paper presents a highly reliable and complete MIR-based colour
selection of AGN with {\rm WISE}. Our technique is designed to select
objects with red MIR power-law SEDs and properly accounts for the
estimated typical errors in the photometry and deviations of the MIR
SEDs from a pure power-law. The AGN wedge is defined using the
wide-angle Bright Ultra-hard \xmm Survey ({\tt BUXS}; Mateos et
al.\ 2012c, in preparation). This survey is one of the largest
complete flux-limited samples of bright ``ultra-hard'' (4.5-10 keV)
X-ray selected AGN to date. Surveys such as \buxs are extremely
efficient in selecting AGN bright enough for reliable optical
identifications and for detailed studies of their properties and
evolution (e.g. HBS28, \citealt{caccianiga04}; HBSS,
\citealt{ceca08}). \buxs covers the region of the AGN
redshift-luminosity parameter space that \wise will sample. Thus,
\buxs offers a unique opportunity to define a highly complete and
reliable MIR-based AGN selection with {\rm WISE}. Thanks to the
optical spectroscopic identifications available for $\sim$97\% of the
\buxs objects, and the high quality X-ray spectra, we have maximized
the completeness of our MIR selection without compromising its
reliability. In a forthcoming paper we will present and discuss the
main properties of the optical/near-IR/MIR SEDs of the AGN in \buxs
(Mateos et al.\ 2012c, in preparation).

This paper is organized as follows. Sections 2 and 3 briefly summarize
the data sets. In Section 4 we present our MIR selection of AGN
candidates using the three shorter wavelength bands of \wise and the
complete four bands, respectively and we discuss the completeness of
the selection. We show the reliability of our AGN selection in Section
5. The results are summarized in Section 6. Throughout this paper
errors are 90\% confidence for a single parameter and we assume
$\Omega_M=0.3$, $\Omega_\Lambda=0.7$ and $H_0={\rm
  70\,km\,s^{-1}\,Mpc^{-1}}$.

\section{The \wise Infrared Survey}
\label{wise_survey}
\wise observed the entire sky in the MIR, achieving 5\sgm point source
sensitivities better than 0.08, 0.11, 1, and 6 mJy at 3.4, 4.6, 12,
and 22 {$\mu$m}, respectively. The angular resolution is 6.1\asec,
6.4\asec, 6.5\asec, and 12.0\asec (FWHM), respectively, and the
astrometric precision for high signal-to-noise (hereafter \sn) sources
is better than 0.15\asec (Wright et al.\ 2010). We use here the March
2012 publicly available All-Sky Data Release that covers $>$99\% of
the sky and incorporates the best available calibrations and data
reduction algorithms (Cutri et al.\ 2012).

In what follows we compute flux densities in the \wise bands using
profile fitting photometry and the magnitude zero points of the Vega
system: $F_\nu{\rm (iso)}$=309.124 Jy, 171.641 Jy, 30.988 Jy, and
8.346 Jy for 3.4, 4.6, 12, and 22 $\mu$m, respectively. These values
are computed with the flux correction factors that correspond to a
power-law spectrum ($f_\nu\propto\nu^{\alpha}$) with spectral index
$\alpha$=$-$1 presented in \citet{wright10}. We note that using the flux
correction factors that correspond to constant power-law spectra the
difference in the computed flux densities would be less than 0.2\% at
3.4, 4.6, and 22 $\mu$m and $\sim$2\% at 12 $\mu$m. We added a 1.5\%
uncertainty to the catalogued flux errors in all bands to account for
the overall systematic uncertainty from the Vega spectrum in the flux
zeropoints. To account for the existing discrepancy between the red
and blue calibrators used for the conversion from magnitudes to
Janskys, an additional 10\% uncertainty was added to the 12\mic and
22\mic fluxes (Wright et al.\ 2010). Throughout this paper we use
monochromatic MIR flux densities ($f_\nu$) in Janskys, unless
otherwise specified.

\begin{figure}
  \centering
  \begin{tabular}{cc}
    \hspace{-0.7cm}\includegraphics[angle=90,width=0.49\textwidth]{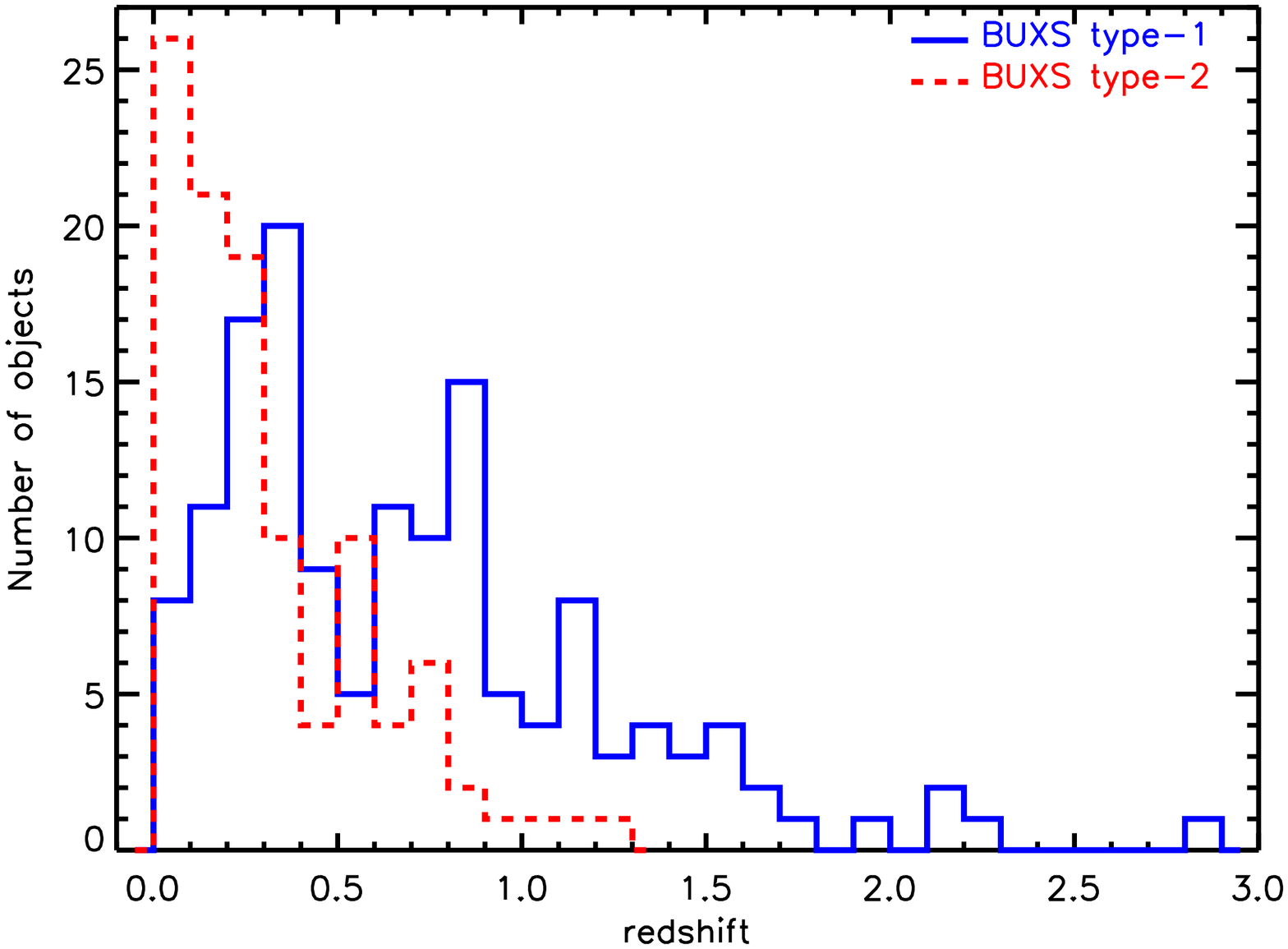}\\
    \hspace*{-0.7cm}\includegraphics[angle=90,width=0.49\textwidth]{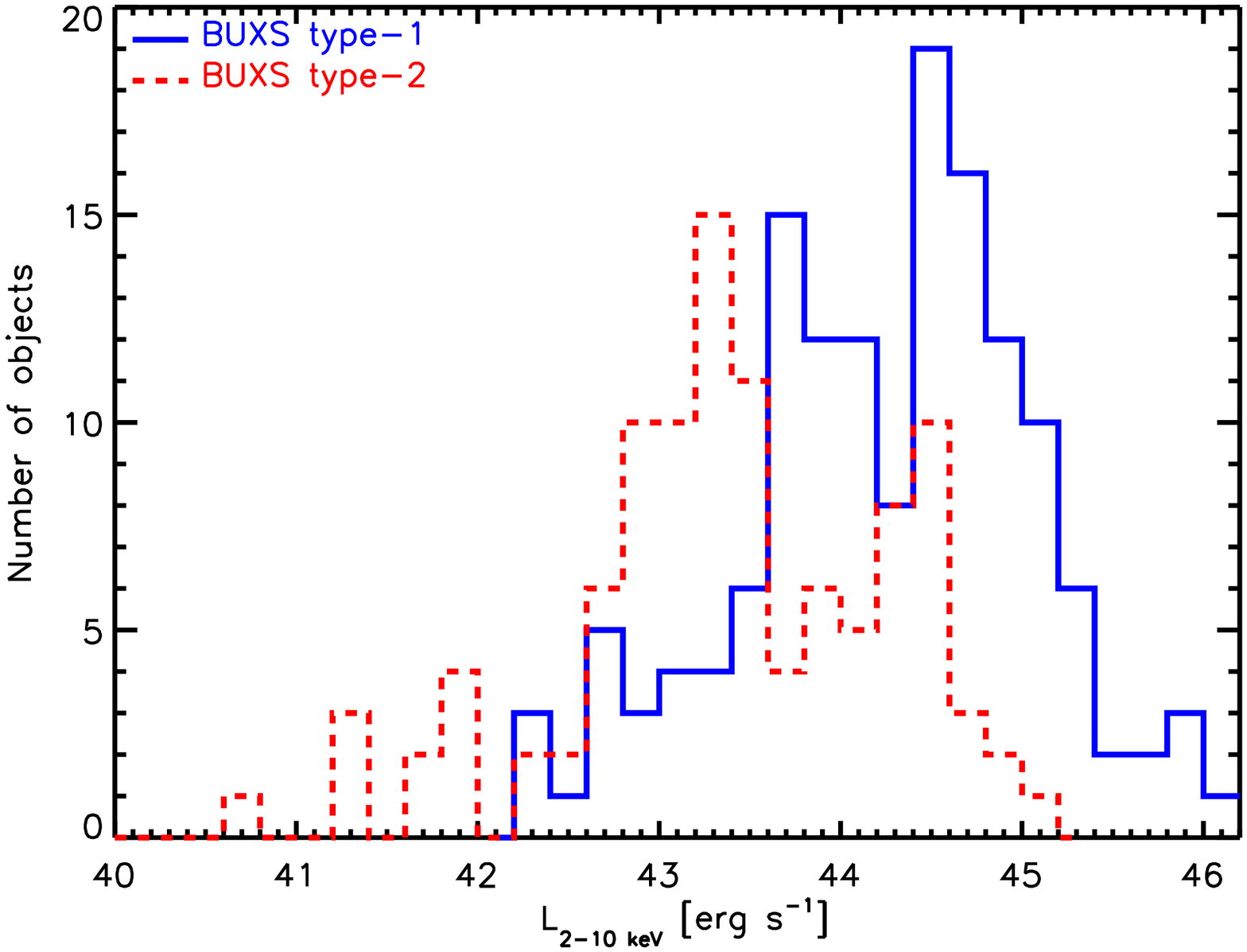}\\
  \end{tabular}
  \caption{Redshift (top) and 2-10 keV luminosity (in log units,
    bottom) distributions of the \buxs type-1 and type-2
    AGN. Luminosities are rest-frame and are corrected for intrinsic
    absorption.}
  \label{fig1}
\end{figure}

\section{The Bright Ultra-hard \xmm Survey}
{\tt BUXS} is one of the largest, amongst the existing \xmm and {\it
  Chandra} surveys, complete flux-limited samples of bright (${\tt {\it
    f}_{4.5-10\,keV} > 6\,x\,10^{-14} \flux}$) ``ultra-hard'' (4.5-10
keV) X-ray selected sources to date.  {\tt BUXS} is based on a subset
of 381 high Galactic latitude ($|b|>20\deg$) observations from the
second \xmm serendipitous source catalogue ({\tt 2XMM};
\citealt{watson09}). The sample is drawn from EPIC-pn observations
with clean exposure times $>$10 ks and having good quality for
serendipitous source detection (i.e. free of bright and/or extended
X-ray sources). These observations were used to derive extragalactic
source count distributions at intermediate fluxes, and therefore we have a
good knowledge of the survey completeness \citep{mateos08}. The
total sky area of {\tt BUXS} is 44.43 deg$^2$.

The selection of sources in the 4.5-10 keV energy band was motivated
by the need to reduce the strong bias against heavily absorbed AGN
affecting surveys conducted at softer energies. The \buxs bright flux
limit was intended to include only objects for which a reliable
classification and redshift could be derived from optical
spectroscopy.  This also ensures that we can derive accurate X-ray
properties such as absorption and intrinsic luminosities from
high-quality X-ray spectra ($>$ few hundred counts).  In this way we
remove uncertainties associated with photometric redshifts and poor
X-ray data quality ($<$ a hundred counts).  {\tt BUXS} contains 258
sources after removal of Galactic stars ($<$2\%) and known BL Lacs
($\sim$1\%).  Optical spectroscopic identifications have been obtained
from the Sloan Digital Sky Survey \citep{abazajian09}, the literature,
and our ongoing follow-up campaign.  At the time of writing, the
spectroscopic identification completeness is 97.3\%. Of the 258 \buxs
sources, 145 objects (56.2\%) are identified as type-1 AGN (UV/optical
emission line velocity widths $\geq$1500 $\kms$) and 106 (41.1\%) as
type-2 AGN (UV/optical emission line velocity widths $<$1500 $\kms$ or
no emission lines). Seven sources (2.7\%) remain unidentified. \buxs
covers four decades in X-ray luminosity (${\rm
  \sim10^{42}-10^{46}\lum}$), where the luminosities are computed in
the `standard' 2-10 keV rest-frame energy band and are corrected for
intrinsic absorption. \buxs identifies sources out to
  \red$\sim$2. Type-1 and type-2 AGN have mean 2-10 keV luminosities
of ${\rm 1.5\times10^{44}\lum}$ and ${\rm 2.2\times10^{43}\lum}$, and
mean redshifts of 0.7 and 0.3, respectively. Redshift and luminosity
distributions are shown in Fig.~\ref{fig1}.  We note that \buxs
samples absorbed AGN in the Compton-thin regime (rest-frame intrinsic
absorption ${\rm N_H\lesssim10^{24}\nh}$).

To find the MIR counterparts of the sources in \buxs we used the
cross-matching algorithm of \citet{pineau11}. The algorithm, which is
based on the classical likelihood ratio, computes the probability of a
spatial coincidence of the X-ray sources with their MIR candidate
counterparts. MIR counterparts were found for 255 out of 258 (98.8\%)
sources (detection with \sn$\geq$5 in at least one of the \wise
bands). The mean X-ray--MIR separation is $\lesssim$2 arcsec. The
three objects without detection in the MIR are type-1 AGN with
\red$\sim$0.6-0.8. For one of these sources a blend of two \wise
objects prevents us from identifying a unique MIR counterpart. For the
other two sources, the expected MIR fluxes, from analysis of their
optical/near-IR SEDs, suggest that they are too faint to be detected
with {\rm WISE}.

In order to assess the completeness of our MIR selection technique with
respect to the overall AGN population we built a clean AGN parent
sample by extracting all catalogued \wise sources in the \buxs survey
area, and we identified the objects detected in the 2-10 keV band
again using the cross-matching algorithm of \citet{pineau11}. Here we
assume that a detection in hard X-rays is a good tracer of unabsorbed
and mildly absorbed AGN activity.  We used the 2-10 keV source lists
from \citet{mateos08} that were derived with the source detection
pipeline used in the {\tt 2XMM} catalogue. The total number of objects
in the \buxs area detected in the 2-10 keV X-ray band is 10265. The
faintest sources have 2-10 keV fluxes of ${\rm
  \sim5\times10^{-15}\flux}$.

We estimate a fraction of spurious matches of X-ray and MIR sources of
$<$1\% from the cross-matching of X-ray and \wise sources using a
large offset in MIR coordinates (3 arcmin in either RA or dec).

\begin{table}
  \begin{center}
    \caption{Summary of the MIR selection of AGN candidates in the \buxs survey area.}
    \label{tab2}
    \begin{tabular}{cccccc}
      \hline
      \hline
      MIR wedge &  ${\rm N_{WISE}}$  &  ${\rm N_{WISE+X}}$ & ${\rm N_{wedge}}$ & ${\rm N_{wedge+X}}$ \\
      (1)       &      (2)         &       (3)         &        (4)          &     (5)               \\  
      \hline
      3-band &     25206  &    1659 (6.6$\%$) &   2755   &    1062 (38.5$\%$)\\
      4-band &     2476   &    409 (16.5$\%$) &   516    &     245 (47.5$\%$)\\
      \hline
    \end{tabular}\\
    Column 1: MIR AGN selection wedge identifier; Column 2: number of
    catalogued \wise sources in the \buxs survey area with
    significance of detection $\geq$5 in the relevant bands; Column
    3: number (fraction) of \wise sources with an X-ray
    detection in the 2-10 keV band; Column 4: number of \wise sources
    in AGN wedge; Column 5: number (fraction) of \wise
    sources in AGN wedge and detected in X-rays.
  \end{center}
\end{table}

\begin{table}
  \begin{center}
    \caption{Summary of the MIR selection of \buxs AGN.}
    \label{tab1}
    \begin{tabular}{ccccccccccc}
      \hline
      \hline
      MIR wedge &  Opt. class  &${\rm N_{WISE}}$  &  ${\rm N_{wedge}}$ \\
      (1)       &      (2)         &       (3)         &        (4) \\
      \hline
      \multirow{3}{*}{3-band} &    Type-1  &    114 &   105\\
                              &    Type-2  &     81 &   38\\
                              &    No ID  &     4  &    3\\
      \hline
      \multirow{3}{*}{4-band} &    Type-1  &     63 &  55 \\
                              &    Type-2  &     55 &  21 \\
                              &    No ID  &      2 &   1 \\
      \hline
    \end{tabular}\\
    Column 1: MIR AGN selection wedge identifier; Column 2: Optical
    class; Column 3: number of sources that are used to define the AGN
    wedge; Column 4: number of objects in AGN wedge.
  \end{center}
\end{table}

\begin{figure*}
  \centering
  \begin{tabular}{cc}
    \hspace{0.0cm}\includegraphics[angle=90,width=0.63\textwidth]{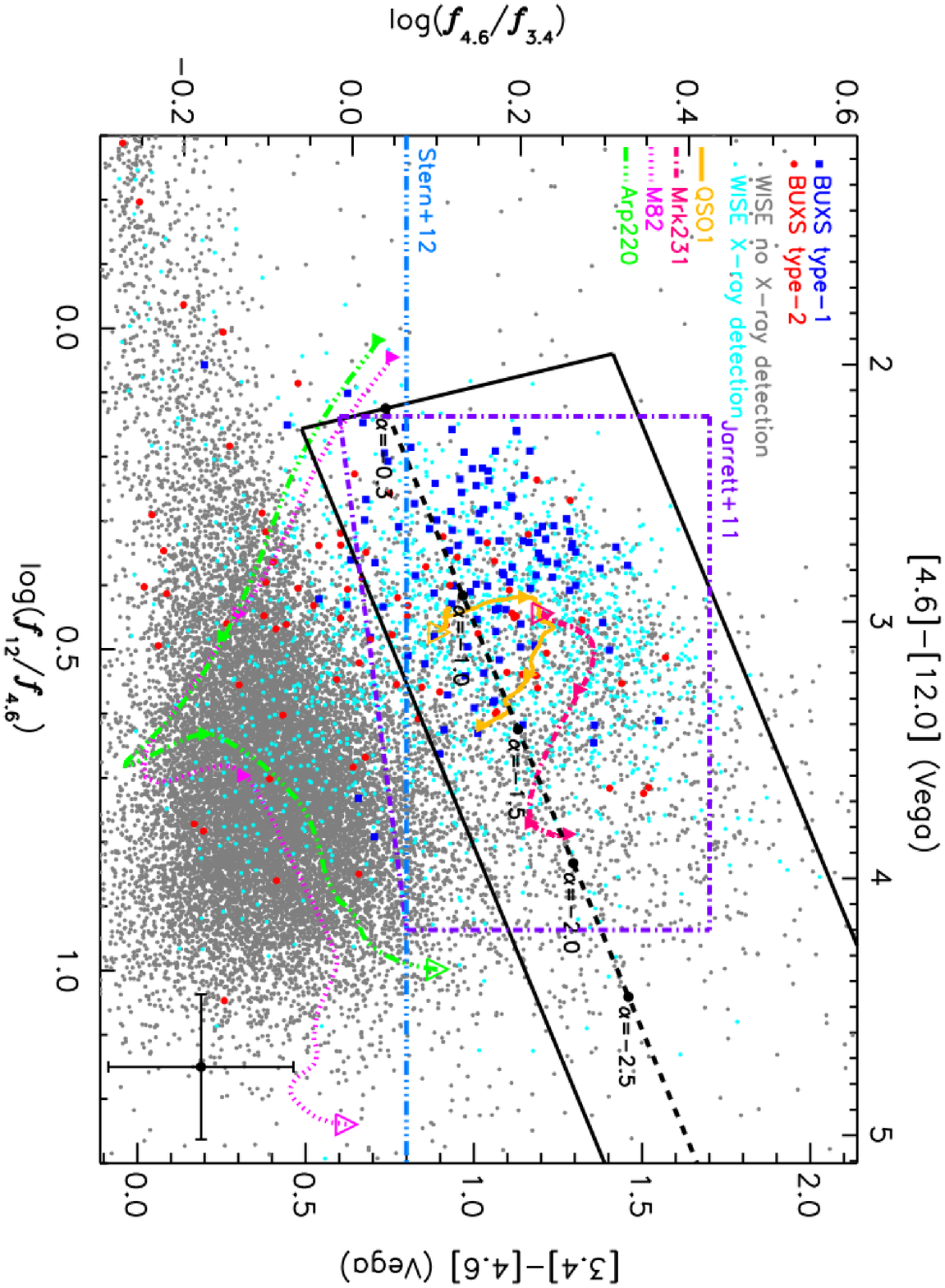}\\
    \vspace*{0.1cm}\\ \includegraphics[angle=90,width=0.55\textwidth]{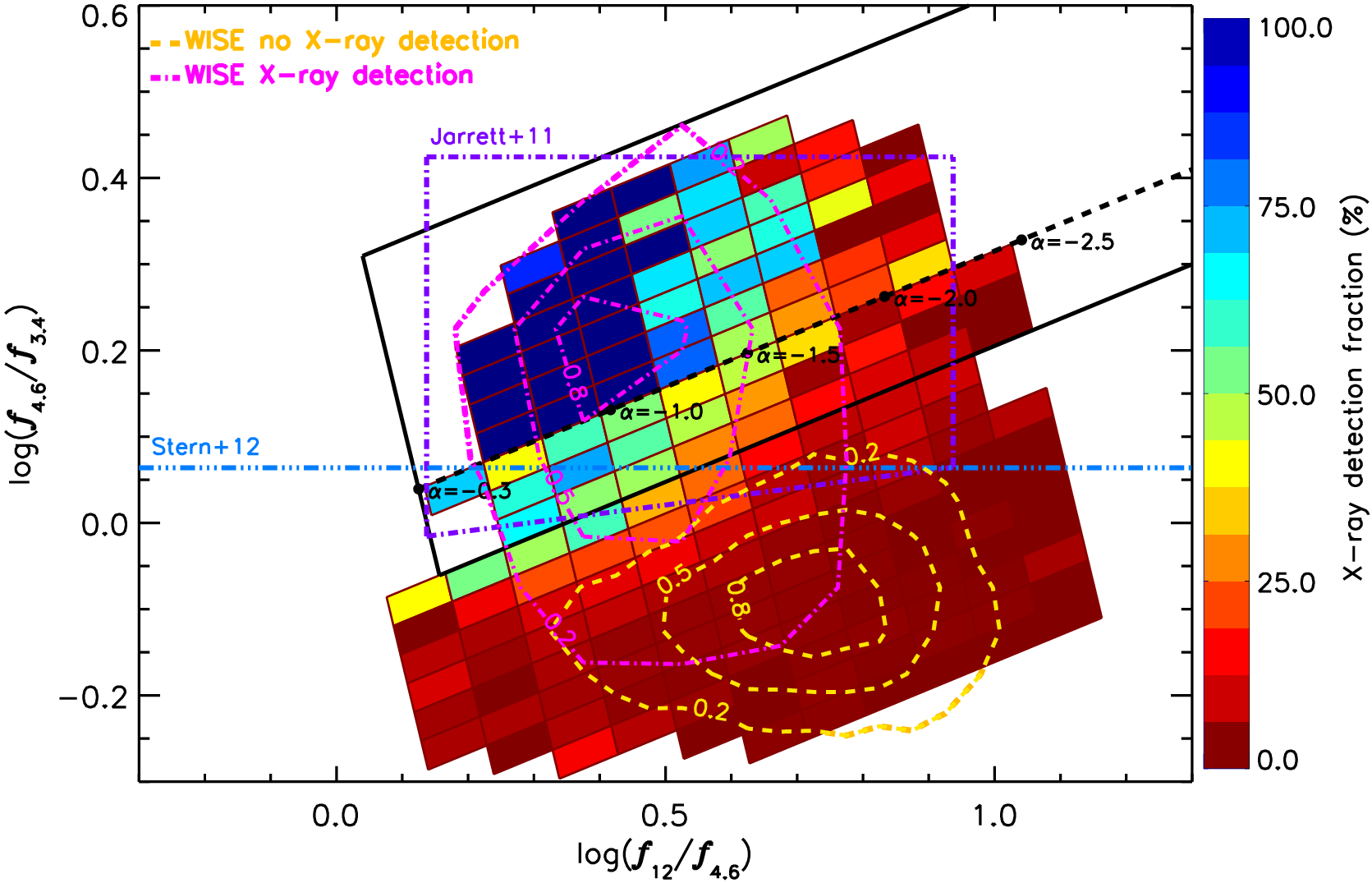}\\
  \end{tabular}
\vspace{+0.2in}
  \caption{Top: MIR colours for sources detected with \sn$\geq$5 at
    3.4, 4.6, and 12 {$\mu$m}. Large symbols represent
    spectroscopically identified {\tt BUXS} AGN. Small cyan and grey
    symbols are \wise sources in the \buxs survey area with and
    without an X-ray detection at 2-10 keV, respectively. The
    0$<$\red$\leq$1.5 ($\Delta$\red=0.5) star-forming tracks represent
    M82 and the ULIRG Arp220. The 0$<$\red$\leq$3 ($\Delta$\red=0.5)
    AGN tracks represent the infrared luminous AGN Mrk231
    \citep{polletta08} and a QSO1 template obtained by stitching
    together the 0.58-3.5\mic near-infrared and 3.5-24\mic MIR quasar
    composite spectrum from \citet{glikman06} and \citet{caballero11}.
    Open symbols indicate \red=0. Our AGN selection wedge and
    power-law locus are the thick solid and dashed black lines,
    respectively. For comparison we show the AGN criteria defined by
    \citet{jarrett11} and \citet{stern12}, respectively (dotted-dashed
    purple and light blue lines). The error bars show the typical
    uncertainties in MIR colours at the \sn=5 limit (see
    Sec.~\ref{3band} for details). Bottom: X-ray detection fraction of
    \wise objects across the colour-colour plane for bins containing
    at least 10 sources. Dotted-dashed (magenta) and dashed (yellow)
    contours indicate the density of \wise sources (normalized to the
    peak value) with and without X-ray detection, respectively.}
  \label{fig2}
\end{figure*}

\section[]{\wise selection of AGN candidates in the \buxs fields}
To avoid spurious detections and objects with poorly constrained
photometry, in what follows we restrict ourselves to \wise sources
detected with \sn$\geq$5 in all relevant bands (3.4, 4.6, and 12\mic
in Sec.~\ref{3band}; 3.4, 4.6, 12, and 22\mic in Sec.~\ref{4band}). 

\subsection[]{\wise three-band AGN wedge}
\label{3band}
In the \buxs survey area there are 25206 sources detected with
\sn$\geq$5 in the three shorter wavelength bands of {\rm WISE}, of
which 1659 have X-ray detections (see Table \ref{tab2}). Out of the
latter, 114 are associated with \buxs type-1 AGN and 81 with \buxs
type-2 AGN (see Table \ref{tab1}). Fig.~\ref{fig2} shows the MIR {\tt
  log}(${\tt {\it f}_{4.6}/{\it f}_{3.4}}$) vs. {\tt log}(${\tt {\it
    f}_{12}/{\it f}_{4.6}}$) diagram for \wise objects with and
without an X-ray counterpart as small cyan and grey symbols,
respectively. We also marked with large blue and red symbols
spectroscopically classified type-1 and type-2 AGN in the \buxs
survey. The dashed line illustrates the MIR power-law locus and the
values for different spectral indices. Most \buxs objects, especially
type-1 AGN, are clustered near the power-law locus, in a region in the
MIR colour-colour plane well separated from the stellar locus (colours
near zero magnitude) and the horizontal sequence of normal galaxies
(lower right part of the diagram). Furthermore, AGN have, on average,
redder {\tt log}(${\tt {\it f}_{4.6}/{\it f}_{3.4}}$) colours than a
pure power-law. This suggests some curvature in the observed 3.4\mic
to 12\mic SEDs (e.g. \citealt{elvis94}; \citealt{richards06};
\citealt{assef10}).

\begin{figure*}
  \centering
  \begin{tabular}{cc}
    \hspace{0.0cm}\includegraphics[angle=90,width=0.63\textwidth]{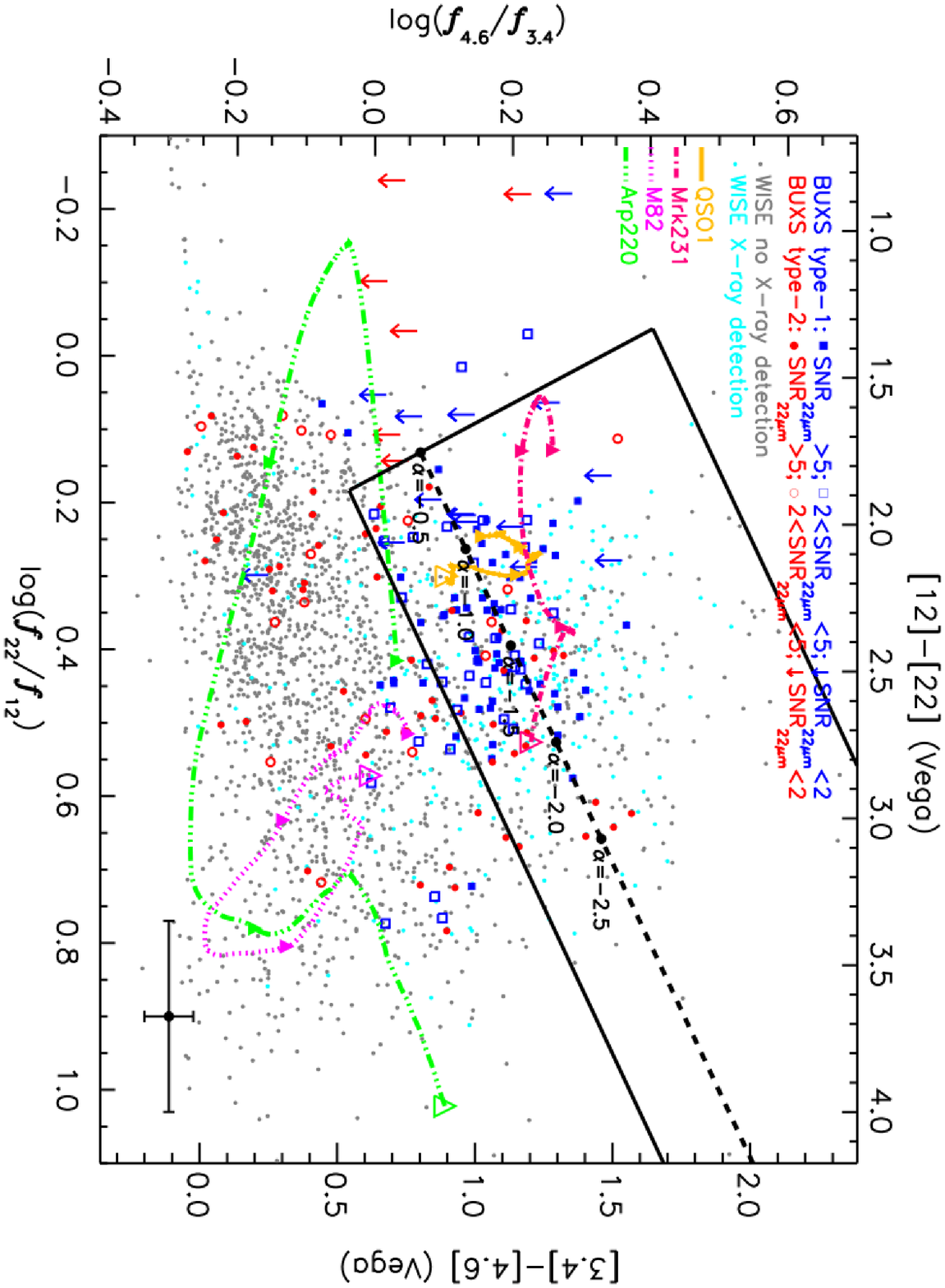}\\
    \vspace{+0.1cm}\\\includegraphics[angle=90,width=0.55\textwidth]{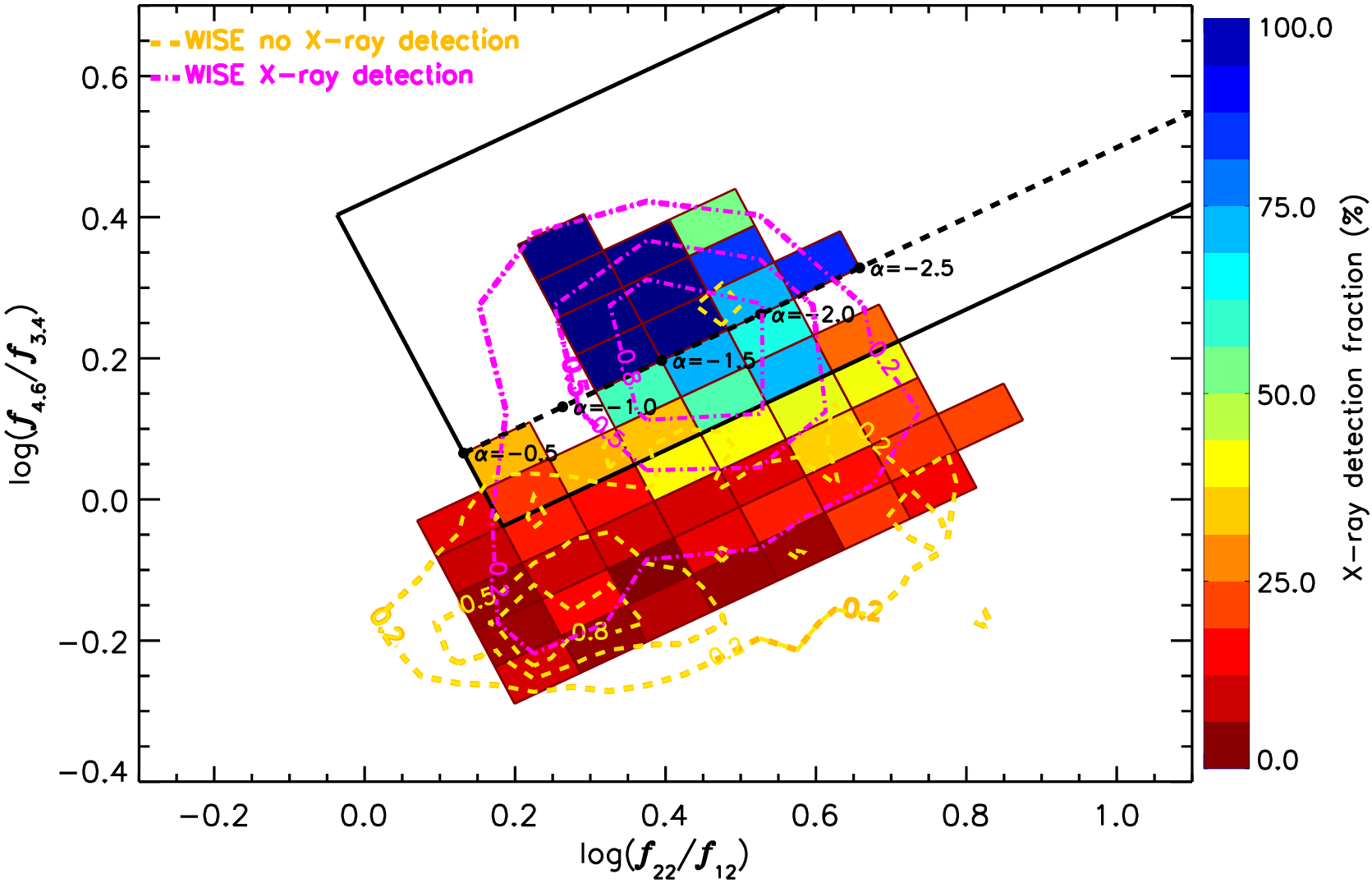}\\
  \end{tabular}
\vspace{+0.2in}
  \caption{Equivalent to Fig.~\ref{fig2} for a MIR-based AGN selection
    using the complete four bands of {\rm WISE}. Top: We include all
    objects with \sn$\geq$5 at 3.4, 4.6, and 12\mic and use different
    symbols depending on the 22\mic significance of detection. For
    \sn$<$2 at 22\mic we use 1$\sigma$ fluxes as upper limits. Symbols as in
    Fig.~\ref{fig2}.}
  \label{fig5}
\end{figure*}

Fig.~\ref{fig2} (bottom) shows the fraction of \wise sources detected
in X-rays across the colour-colour diagram and the distribution of
objects with and without detection in X-rays (contours). There is a
clear separation between these distributions. The bulk of the MIR
population not detected in X-rays overlaps with the horizontal
sequence of normal galaxies, while the great majority of X-ray
detected objects cluster near the power-law locus.

Our MIR-based AGN selection technique is designed to identify objects
with red MIR power-law SEDs.  The AGN wedge is defined to include all
objects with MIR colours expected for power-law SEDs with spectral
index $\alpha\leq$$-$0.3, properly accounting for the typical errors in
the photometry at the faintest MIR fluxes (error bars in
Fig.~\ref{fig2}, top). We then increase the size of the wedge towards red
{\tt log}(${\tt {\it f}_{4.6}/{\it f}_{3.4}}$) colours (upper
boundary) to include all \buxs AGN and the X-ray detected \wise
objects throughout the \buxs survey area with such colours. In this
way we account for deviations of the 3.4\mic to 12\mic SEDs from a
pure power-law. Our three-band AGN wedge is shown with the thick solid
box in Fig.~\ref{fig2}. The MIR power-law locus is defined by
\begin{eqnarray}
\hspace*{0.1cm} y=0.315 \times x 
\end{eqnarray}
where $x\equiv\rm{log_{10}}\left(\frac{{\it f}_{12\mu m}}{{\it
    f}_{4.6\mu m}}\right)$ and $y\equiv\rm{log_{10}}\left(\frac{{\it
    f}_{4.6\mu m}}{{\it f}_{3.4\mu m}}\right)$.
The top and bottom boundaries of the wedge are obtained by adding
y-axis intercepts of +0.297 and $-$0.110, respectively. The MIR power-law $\alpha$=$-$0.3
bottom-left limit corresponds to 
\begin{eqnarray}
\hspace*{0.1cm} y=-3.172 \times x+0.436 
\end{eqnarray}
These are the limits we have used throughout this paper. However, for most
practical purposes using instead a bottom-left vertical limit, corresponding to
$x$$\ge$0.120, produces very similar results.

In Vega magnitudes (in the \wise source catalogue magnitudes are
reported in the Vega system), the AGN locus is defined by 
\begin{eqnarray}
\hspace*{0.1cm} y'=0.315 \times x' 
\end{eqnarray}
where $x$'$\equiv[4.6]-[12]$ and $y$'$\equiv[3.4]-[4.6]$. 
The top and bottom boundaries of the wedge are obtained by adding y-axis intercepts of
+0.796 and $-$0.222, respectively. In this case, the MIR power-law $\alpha$=$-$0.3
bottom-left limit corresponds to 
\begin{eqnarray}
\hspace*{0.1cm} y'=-3.172 \times x'+7.624
\end{eqnarray}
The bottom-left vertical limit of the AGN wedge in magnitudes
corresponds to $x$'$\ge$2.157.\\

Our three-band AGN wedge identifies 2755 AGN candidates in the \buxs
area, of which 1062 (38.5\%) are detected in X-rays (see Table
\ref{tab2}). Out of the latter, 105 are associated with \buxs type-1
AGN and 38 with \buxs type-2 AGN (see Table \ref{tab1}). We note
  that the X-ray detection fraction in the wedge increases with the
  depth of the X-ray observations as shown in Table \ref{tab6}. For
  example, in the \buxs area where the X-ray observations have
  exposures $>$40\,ks, the X-ray detection fraction rises to
  49.8\%. For comparison, in the 1Ms CDF-S survey the X-ray detection
  fraction of IRAC power-law AGN candidates was $\sim$50\%
  \citep{alonso06} while this fraction increased to $\sim$85\% in the
  deeper 2Ms CDF-N survey \citep{donley07}. Furthermore,
  \citet{donley12} found that the X-ray detection fraction of IRAC MIR
  AGN candidates in COSMOS increased from 38\% to 52\% in the regions
  of deep \chandra coverage (X-ray exposures 50-160 ks). This is as
  expected, as long X-ray exposures are required to detect
  intrinsically less luminous and/or heavily obscured AGN (see
  e.g. \citealt{mateos05}; \citealt{tozzi06}; \citealt{comastri11};
  \citealt{brightman12}). Still, a substantial fraction of our MIR AGN
  candidates are undetected at 2-10 keV energies with the typical
  exposures in the {\tt 2XMM} catalogue. These objects that have the
  reddest overall {\tt log}(${\tt {\it f}_{12}/{\it f}_{4.6}}$)
  colours in the AGN wedge, are the best candidates to account for the
  most heavily obscured/absorbed luminous AGN missed by hard X-ray
  surveys. \citet{lacy07} presented the optical spectroscopic followup
  of a sample of luminous AGN candidates selected on the basis of
  their IRAC MIR colours. They confirmed the AGN nature of 91\% of the
  sources, with the majority of the objects being identified as
  dust-reddened type-1 quasars and type-2 AGN. Furthermore, a detailed
  study of the \wise MIR SEDs of [OIII]5007\AA-selected QSO2s
  \citep{reyes08} strongly supports our hypothesis that many of the
  X-ray undetected sources in the wedge are heavily-obscured
  very-luminous AGN (Mateos et al. 2012b, in prep.). 

\begin{table}
  \begin{center}
    \caption{Dependence of the X-ray detection fraction of \wise
      sources in the 3-band AGN wedge as a function of the exposure time of the
      \xmm observations.}
    \label{tab6}
    \begin{tabular}{ccccc}
      \hline
      \hline
       ${\rm t_{exp}}$ & ${\rm N_{wedge}}$ & ${\rm N_{wedge+X}}$  \\
      (1) & (2) & (3)  \\
      \hline
      10$-$20           & 1225   &  400 (32.7\%)                      \\  
      20$-$30           &  723   &  298 (41.2\%)                      \\  
      30$-$40           &  347   &  135 (38.9\%)                      \\  
      40$-$50           &  210   &  110 (52.4\%)                      \\  
      $>$50             &  250   &  119 (47.6\%)                      \\  
      \hline
      Total   & 2755     &   1062 (38.5$\%$)   \\
      \hline
    \end{tabular} \\
    Column 1: EPIC-pn exposure time interval of the X-ray observations
    in units of ks; Column 2: number of catalogued \wise sources in
    AGN wedge with significance of detection $\geq$5 in the three
    shorter wavelength bands of $\rm WISE$. Column 3: number
    (fraction) of \wise sources in AGN wedge with an X-ray detection
    in the 2-10 keV band.
  \end{center}
\end{table}

\subsection[]{\wise four-band AGN wedge}
\label{4band}
Due to the significantly shallower depth at 22\mic compared with the
first three bands, a selection that uses the 22\mic survey will be
restricted by necessity to the brightest MIR objects. We have
investigated, however, whether we can gain any additional information
on AGN selection by using the complete four \wise bands.  The number
of MIR sources detected with \sn$\geq$5 in all four \wise bands in the
area of \buxs is 2476, of which 409 are detected in X-rays. Out of the
latter, 63 are associated with \buxs type-1 AGN and 55 with \buxs
type-2 AGN (see Tables \ref{tab2} and \ref{tab1}). Fig.~\ref{fig5}
(top) shows the distribution of {\tt log}(${\tt {\it f}_{4.6}/{\it
    f}_{3.4}}$) vs. {\tt log}(${\tt {\it f}_{22}/{\it f}_{12}}$)
colours. The solid lines illustrate the four-band AGN selection wedge as
in Sec.~4.1.  In this case we increase the size of the wedge that
would be required to account for the typical photometric errors (error
bars in Fig.~\ref{fig5}) towards both redder {\tt log}(${\tt {\it
    f}_{22}/{\it f}_{12}}$) and {\tt log}(${\tt {\it f}_{4.6}/{\it
    f}_{3.4}}$) colours. In this way we increase the completeness of
the selection without compromising the reliability. The MIR power-law locus is defined by
\begin{eqnarray}
\hspace*{0.1cm} y=0.50 \times x 
\end{eqnarray}
where $x\equiv\rm{log_{10}}\left(\frac{{\it f}_{22\mu m}}{{\it
    f}_{12\mu m}}\right)$ and $y\equiv\rm{log_{10}}\left(\frac{{\it
    f}_{4.6\mu m}}{{\it f}_{3.4\mu m}}\right)$.
The top and bottom boundaries of the wedge are obtained by adding y-axis intercepts of
+0.421 and $-$0.130, respectively. In this case we use a MIR power-law 
bottom-left limit of $\alpha$=$-$0.5 that corresponds to 
\begin{eqnarray}
\hspace*{0.1cm} y=-2.00 \times x+0.33
\end{eqnarray}
These are the limits we have used in Sec.~\ref{completeness}. Using instead a
bottom-left vertical limit, corresponding to $x$$\ge$0.13, produces
very similar results.

In Vega magnitudes the AGN locus is defined by
\begin{eqnarray}
\hspace*{0.1cm} y'=0.50 \times x' 
\end{eqnarray}
where $x$'$\equiv[12]-[22]$ and $y$'$\equiv[3.4]-[4.6]$. The top and
bottom boundaries of the wedge are obtained by adding y-axis
intercepts of +0.979 and $-$0.405, respectively. In this case,
  the MIR power-law $\alpha$=$-$0.5 bottom-left limit corresponds to
\begin{eqnarray}
\hspace*{0.1cm} y'=-2.00 \times x'+4.33
\end{eqnarray}
The bottom-left vertical limit of the AGN wedge in magnitudes
corresponds to $x$'$\ge$1.76.
\\

Our four-band AGN wedge identifies 516 AGN candidates in the \buxs
area detected with \sn$\geq$5 in all four \wise bands, of which 245
(47.5\%) are detected in X-rays. Out of the latter, 55 are associated
with \buxs type-1 AGN and 21 with \buxs type-2 AGN (see Table
\ref{tab1}).

\begin{figure}
  \centering
  \begin{tabular}{cc}
    \hspace{-0.7cm}\includegraphics[angle=90,width=0.49\textwidth]{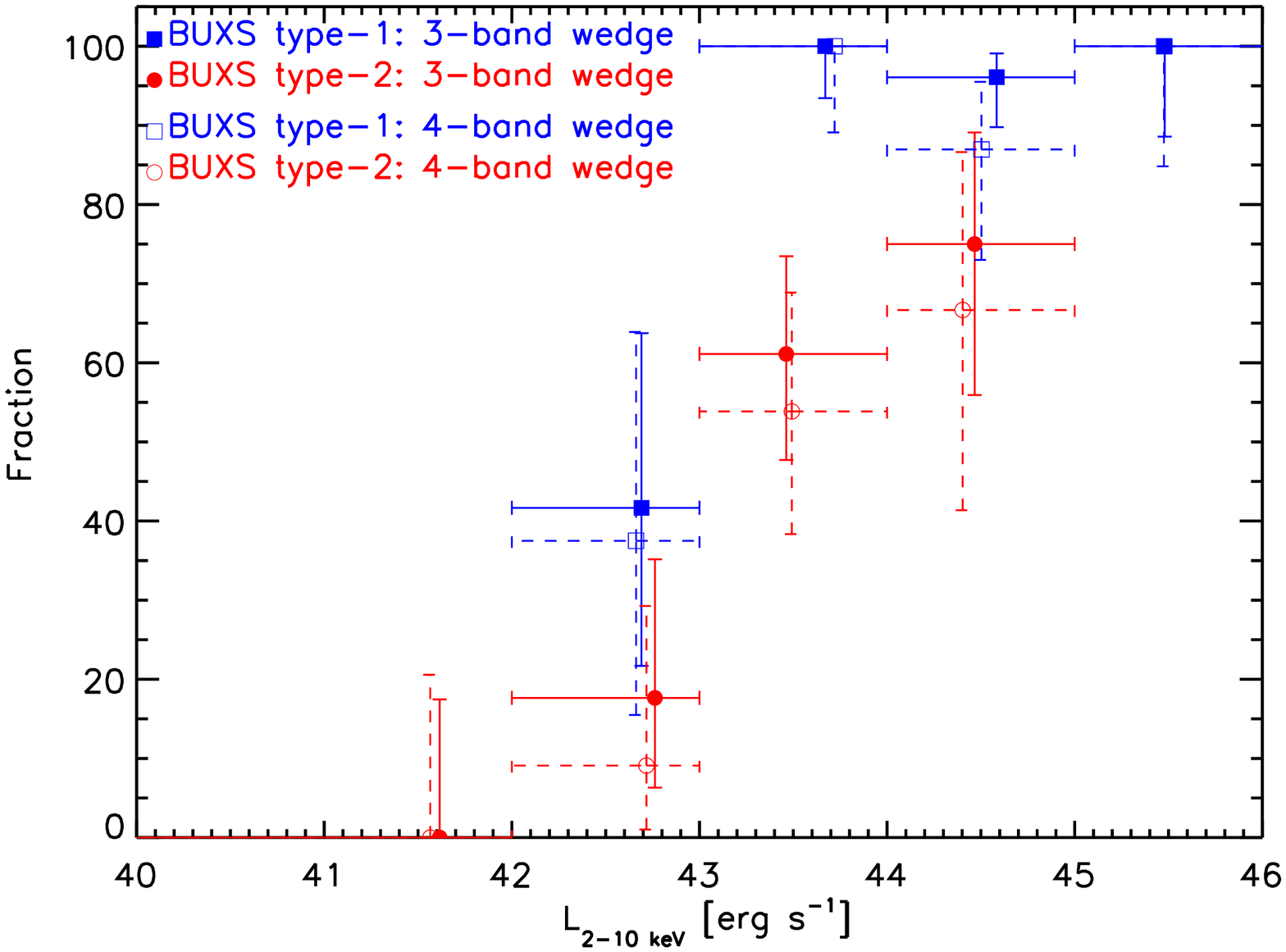}\\
    \hspace{-0.7cm}\includegraphics[angle=90,width=0.49\textwidth]{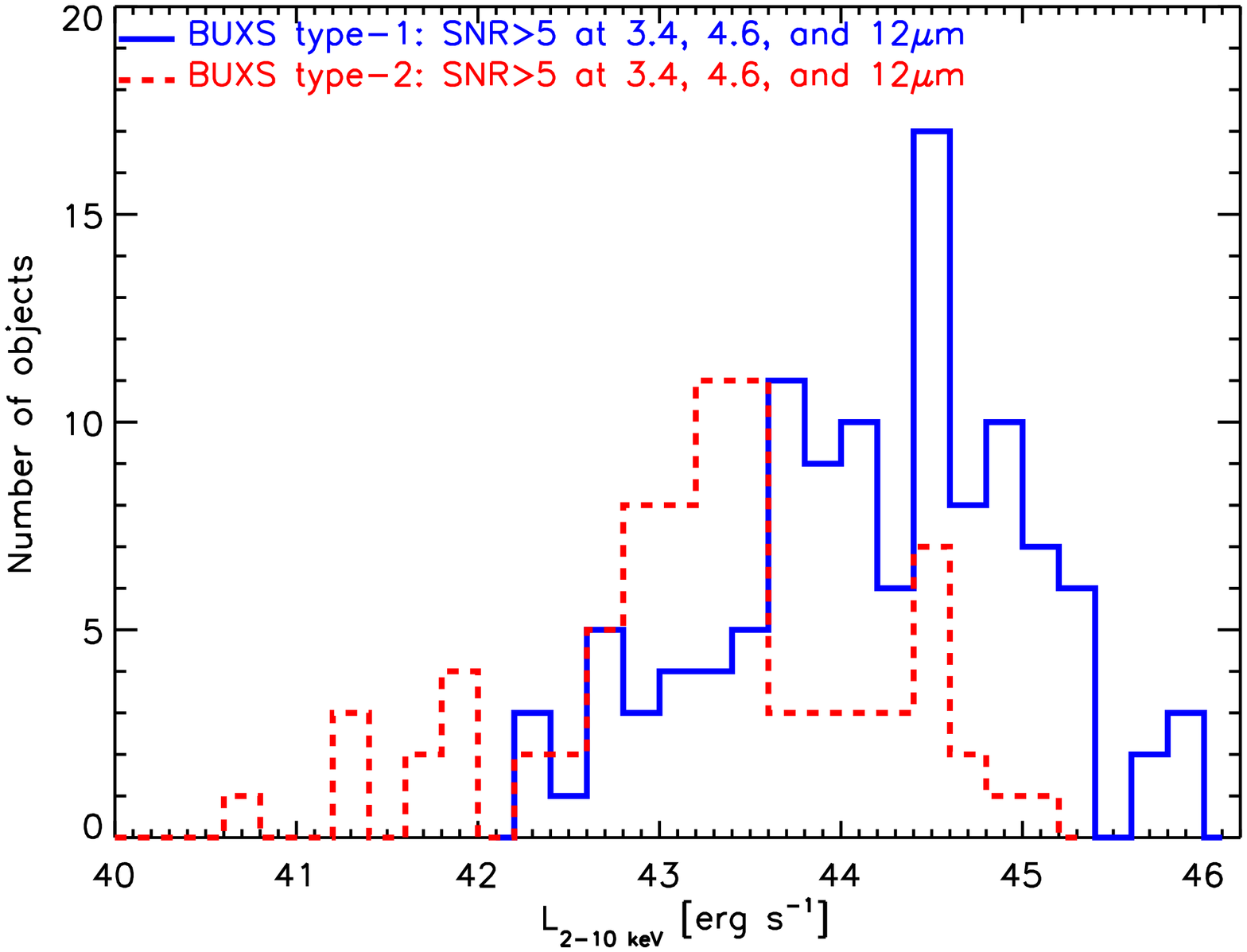}\\
  \end{tabular}
  \caption{Top: Fraction of \buxs AGN that meet our MIR selection as a
    function of their intrinsic 2-10 keV luminosity (in log
    units). Filled symbols show selection using the three shorter
    wavelength bands of \wise and open symbols show selection using
    the complete four bands. The symbols indicate the mean luminosity
    of the sources in the bin. Bottom: Distributions of 2-10 keV
    luminosity (in log units) for the \buxs type-1 and type-2 AGN
    detected with \sn$\geq$5 in the three shorter wavelength bands of
    {\rm WISE}. The luminosity (and redshift)
    distributions of the \buxs AGN detected in the three shorter
    wavelength bands of \wise and the complete four bands are
    consistent with each other.}
  \label{fig3}
\end{figure}

\subsection[]{AGN selection completeness}
\label{completeness}
Fig.~\ref{fig3} (top) and Table \ref{tab3} show the fraction of \buxs
AGN that meet our three-band MIR colour cuts as a function of their
intrinsic 2-10 keV luminosity (filled symbols). The symbols indicate
the mean luminosity of the sources in the bin. Rest-frame luminosities
were derived from a detailed X-ray spectroscopic analysis and are
corrected for Galactic and any intrinsic absorption. The completeness
of our selection criteria is a strong function of luminosity for both
type-1 and type-2 AGN. This result reflects the fact that objects with
MIR colours not dominated by the thermal emission from the AGN will be
missed by our selection. This effect is more important for
low-luminosity AGN, especially if these sources are affected by large
dust extinction at the shortest wavelengths of {\rm WISE}. In these
objects the starlight from the host galaxy will dominate their MIR
emission. Thus, it should be emphasized that the comparison of an MIR
colour selection completeness for different classes of objects is only
meaningful if the objects span the same range of luminosities. Taking
this into account, at ${\tt {\it L}_{2-10\,keV}<10^{44}\lum}$,
$84.4_{-10.0}^{+7.4}\%$ and $39.1_{-9.5}^{+10.1}\%$ of the type-1 and
type-2 AGN, respectively meet the selection. At ${\tt {\it
    L}_{2-10\,keV}\geq 10^{44}\lum}$ the MIR selection efficiency
increases to $97.1_{-4.8}^{+2.2}\%$ and $76.5_{-18.4}^{+13.3}\%$ for
type-1 and type-2 AGN, respectively. We estimated the most probable
value for these fractions using a Bayesian approach and the binomial
distribution from \citet{wall08}. The quoted errors are the narrowest
interval that includes the mode and encompasses 90\% of the
probability (S. Andreon, private communication). It is important to
note that the significantly smaller value of the selection
completeness obtained for type-2 AGN at ${\tt {\it
    L}_{2-10\,keV}<10^{44}\lum}$ compared to that for type-1 AGN is
mainly due, as indicated above, to the different luminosity
distributions of the two classes of AGN. The type-2 AGN population in
\buxs is dominated by objects with ${\tt {\it L}_{2-10\,keV}\lesssim
  10^{44}\lum}$ ($\sim$79\% vs. $\sim$39\% for type-1 AGN; see
Fig.~\ref{fig3} bottom). At such luminosities many AGN have relatively
blue colours at the shortest \wise wavelengths (i.e. host-dominated)
and thus lie outside the MIR AGN wedge. We expect this effect to be
more important for type-2 AGN, as this class of objects is expected to
show a higher degree of extinction at the shortest wavelengths of {\rm
  WISE}. For example, the clumpy torus models of \citet{nenkova08}
predict nearly isotropic emission at wavelengths
$\gtrsim$12{$\mu$m}. This could explain that even if we use the same
luminosity range (as in Fig.~\ref{fig3}) we obtain a selection
completeness that is, within the uncertainties, still marginally lower
for type-2 AGN than for type-1 AGN. This result still holds at
luminosities ${\tt {\it L}_{2-10\,keV}>10^{44}\lum}$, where the
relative contribution of the host galaxy to the MIR emission should be
small. However, as the number of type-2 AGN in \buxs at such
luminosities is small (16 objects), the difference could be due in
part to small number statistics. Thus, we conclude that our 3-band AGN
wedge is highly complete for both X-ray selected luminous type-1 and
type-2 AGN.

\begin{table}
  \begin{center}
    \caption{Dependence of the \wise 3-band AGN wedge completeness on the luminosity of the \buxs AGN.}
    \label{tab3}
    \begin{tabular}{ccccc}
      \hline
      \hline
      ${\rm log(L_{2-10\,keV})}$ &  ${\rm N_{type-1}}$  &  ${\rm {\it f}_{type-1}}$  &  ${\rm N_{type-2}}$  &  ${\rm {\it f}_{type-2}}$  \\
      (1) & (2) & (3) & (4) & (5) \\
      \hline
      {[}40$-$42{]} &   -  &              -                &    11  &  ${\rm 0.0_{}^{+17.4}}$ \\  
      {[}42$-$43{]} &  12  &    ${\rm 41.7_{-20.0}^{+22.1}}$  &    17  &  ${\rm 17.6_{-11.3}^{+17.5}}$\\   
      {[}43$-$44{]} &  33  &    ${\rm 100_{-6.6}^{}}$       &    36  &  ${\rm 61.1_{-13.4}^{+12.3}}$\\
      {[}44$-$45{]} &  51  &    ${\rm 96.1_{-6.3}^{+3.0}}$   &    16  &  ${\rm 75.0_{-19.1}^{+14.1}}$\\
      {[}45$-$46{]} &  18  &    ${\rm 100_{-11.4}^{}}$      &     1   &      -                   \\ 
      \hline
      Total   & 114     &                             &  81      &                     \\
      \hline
    \end{tabular} \\
Column 1: X-ray luminosity range in units of $\lum$ (logarithmic units,
2-10 keV in rest-frame and corrected for any intrinsic absorption);
Column 2: number of \buxs type-1 AGN in luminosity bin; Column 3:
fraction of \buxs type-1 AGN in the \wise 3-band AGN wedge; Column 4:
number of \buxs type-2 AGN in luminosity bin; Column 5: fraction of
\buxs type-2 AGN in the \wise 3-band AGN wedge.
  \end{center}
\end{table}

\begin{table}
  \begin{center}
    \caption{Dependence of the \wise 4-band AGN wedge completeness on the luminosity of the \buxs AGN.}
    \label{tab4}
    \begin{tabular}{cccccc}
      \hline
      \hline
      ${\rm log(L_{2-10\,keV})}$ &  ${\rm N_{type-1}}$  &  ${\rm {\it f}_{type-1}}$  &  ${\rm N_{type-2}}$  &  ${\rm {\it f}_{type-2}}$  \\
      (1) & (2) & (3) & (4) & (5) \\
      \hline
      {[}40$-$42{]} &   -  &              -                &    9   &  ${\rm 0.0_{}^{+20.6}}$   \\
      {[}42$-$43{]} &  8   &    ${\rm 37.5_{-22.0}^{+26.4}}$  &    11  &  ${\rm 9.1_{-8.1}^{+20.2}}$   \\
      {[}43$-$44{]} &  19  &    ${\rm 100_{-10.9}^{}}$       &    26  &  ${\rm 53.8_{-15.5}^{+15.0}}$   \\
      {[}44$-$45{]} &  23  &    ${\rm 87.0_{-14.0}^{+8.5}}$   &    9   &  ${\rm 66.7_{-25.3}^{+20.0}}$   \\
      {[}45$-$46{]} &  13  &    ${\rm 100_{-15.2}^{}}$      &    -    &   -                           \\
      \hline
      Total   & 63     &                             &  55     &                              \\
      \hline
    \end{tabular} \\
Column 1: X-ray luminosity range in units of $\lum$ (logarithmic
units, 2-10 keV in rest-frame and corrected for any intrinsic
absorption); Column 2: number of \buxs type-1 AGN in luminosity bin;
Column 3: fraction of \buxs type-1 AGN in the \wise 4-band AGN wedge;
Column 4: number of \buxs type-2 AGN in luminosity bin; Column 5:
fraction of \buxs type-2 AGN in the \wise 4-band AGN wedge.
  \end{center}
\end{table}

\begin{figure*}
  \centering
  \begin{tabular}{cc}
    \hspace{-0.0cm}\includegraphics[angle=90,width=0.95\textwidth]{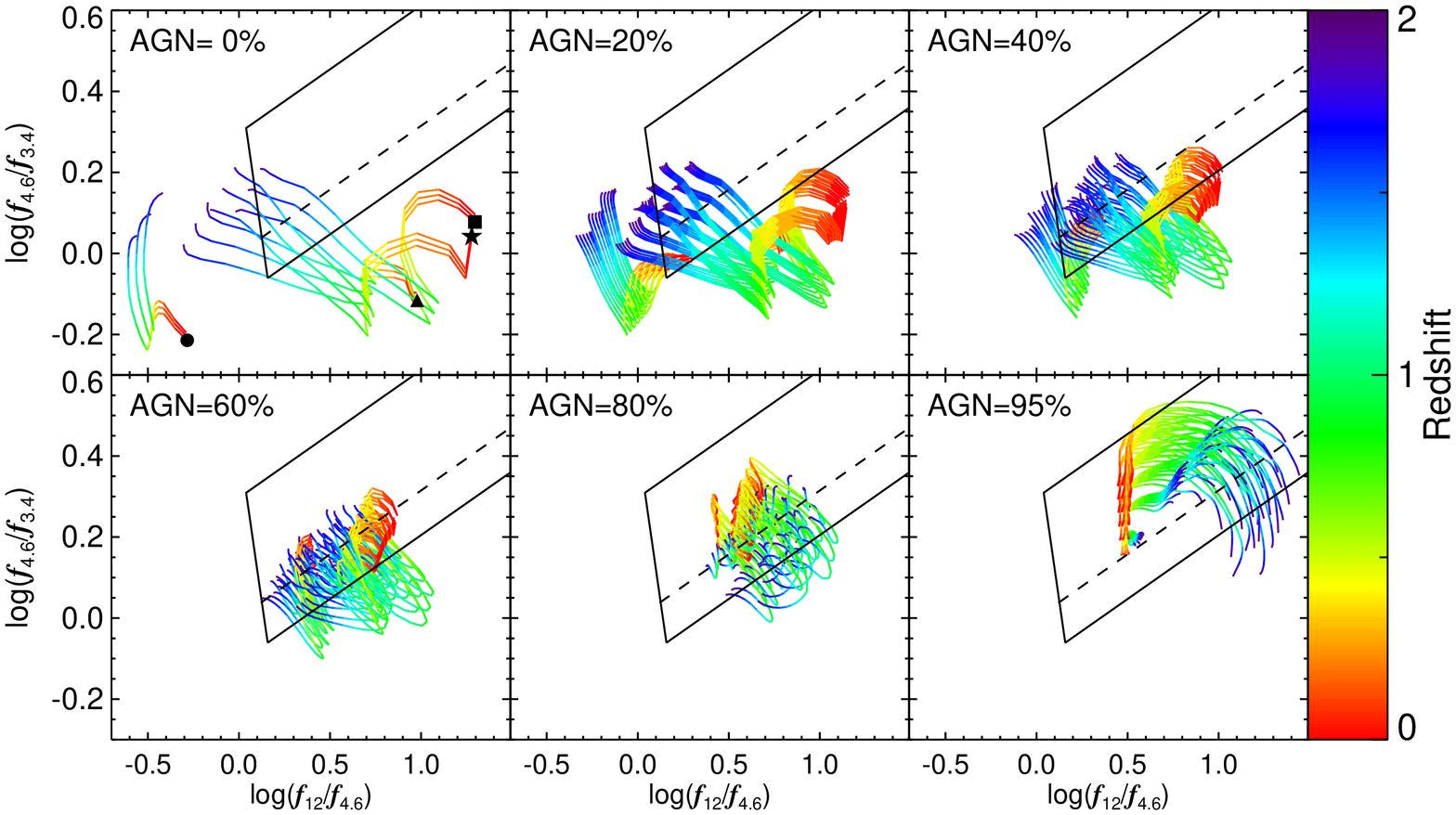}\vspace{-3.0cm}\\
  \end{tabular}
  \vspace{+0.0cm}\caption{Predicted \red=0-2 \wise colours of
    AGN/galaxy composite SEDs for our three-band AGN wedge. The AGN
    fraction is defined between 1 and 10 \micron.  The star-forming
    templates represent the ULIRG IRAS 22491 (square,
    \citealt{polletta08}), the starburst M82 (star,
    \citealt{polletta08}), a normal star-forming spiral galaxy
    (triangle, \citealt{dale02}), and an elliptical galaxy (circle,
    \citealt{polletta08}). Large symbols mark each family of purely
    star-forming templates at \red=0. The AGN template is the QSO1
    template of \citet{polletta08}.  Additional extinctions of $A_{\rm
      V}=0-2$ and $A_{\rm V}=0-20$ are applied to the star-forming and
    AGN components, respectively. Our MIR AGN selection wedge and
    power-law locus are the thick solid and dashed black lines,
    respectively. The power-law locus is defined from $\alpha$=$-$0.3.
    The colour tracks of purely star-forming galaxies would enter our
    selection wedge at \red $\gtrsim$1.3. At such redshifts however,
    most galaxies are too faint to be detected at the relatively
    shallow 12\mic flux density limits.}
  \label{fig6}
\end{figure*}

Fig.~\ref{fig3} (top) and Table \ref{tab4} show the fraction of \buxs
AGN that meet our four-band colour cuts as a function of their
intrinsic 2-10 keV luminosity (open symbols). The completeness of the
four-band wedge is somewhat smaller but comparable, within the
uncertainties, to that achieved with the three-band selection for both
type-1 and type-2 AGN. Indeed, 76 out of the 118 \buxs
AGN detected in the four \wise bands meet the four-band selection,
while this number increases to 88 if the three-band selection is used
instead.  All the 15 three-band selected \buxs AGN that miss the
four-band selection have {\tt log}(${\tt {\it f}_{22}/{\it f}_{12}}$)
colours significantly redder than those expected for a pure power-law
SED. These objects lie outside the four-band AGN wedge, in the region
of the colour-colour plane occupied by normal star-forming
galaxies. They likely miss the four-band selection because their
22\mic emission comes from both AGN activity and intense star
formation. On the other hand, only three \buxs AGN from the four-band
selection lie outside the three-band wedge. We note that we obtain the
same result if we include all \buxs AGN with a lower significance of
detection at 22\mic (open symbols and arrows in Fig.~\ref{fig5},
top). Therefore, by requiring 22\mic detections we are not biased
against AGN with pure power-law SEDs (i.e. the faintest objects at
22{$\mu$m}). Thus, by including the 22{$\mu$m} \wise
band to select AGN candidates neither the completeness nor the
reliability of the selection improves.

We have checked that using non-contiguous bands to define the \wise
MIR colours does not increase the completeness of a three-band or a
four-band selection.

\section[]{Reliability of the three-band AGN wedge}
In the following we restrict ourselves to the three-band AGN wedge
which, as shown in the previous section, provides the most complete
selection of AGN candidates in the \buxs fields with {\rm WISE}.

\subsection[]{Comparison with templates}
A known limitation of MIR selection techniques is the contamination
from galaxies without AGN activity where the major contributor to the
MIR emission is the stellar population or strong star formation
(e.g. \citealt{lacy04}; \citealt{stern05}; \citealt{jarrett11};
\citealt{donley12}; \citealt{stern12}). To assess the reliability of
our three-band AGN wedge we show in Fig.~\ref{fig6} the expected \wise
colours of AGN with varying host-galaxy contributions following
\citet{donley12}. The composite SEDs were constructed using the
library of \citet{polletta08} and the normal star-forming spiral
galaxy template from \citet{dale02}. The solid and dashed lines
illustrate the AGN wedge along with the power-law locus,
respectively. We also applied additional extinctions of ${\rm
  A_V}$=0-2 and ${\rm A_V}$=0-20 to the star-forming and AGN
components, respectively, using the \citet{draine03} extinction
curve. The colour tracks of pure star-forming galaxies would enter our
selection wedge at \red $\gtrsim$1.3.  At such redshifts, however, most
galaxies are too faint to be detected at the relatively shallow \wise
flux density limits (\citealt{wright10}; \citealt{jarrett11}).
Indeed, in the region of the AGN wedge where we would expect
contamination from normal star-forming galaxies, there is no clear
excess of MIR X-ray undetected sources and the X-ray detection
fraction remains high (see Fig.~\ref{fig2}). Thus it seems that at the
adopted \sn$\geq$5 limit our AGN selection suffers from minimal
contamination from high redshift pure star-forming galaxies. It is
interesting to note that an important fraction of the X-ray detected
\wise sources has {\tt log}(${\tt {\it f}_{12}/{\it f}_{4.6})}$
colours bluer than those expected for a pure AGN SED. The \wise colour
tracks of composite galaxies suggest that this is most likely due to
both AGN and their host galaxies contributing to the observed emission
in the \wise bands. On the other hand, we find a sharp decrease in the
X-ray detection fraction of \wise objects at {\tt log}(${\tt {\it
    f}_{12}/{\it f}_{4.6})\gtrsim}$0.7-0.8. At such red MIR colours we
expect many objects to be heavily obscured AGN. However, the colour
tracks of pure AGN indicate that there is a strong dependence of the
observed {\tt log}(${\tt {\it f}_{12}/{\it f}_{4.6})}$ colour with
redshift. This suggests that the population of \wise objects with {\tt
  log}(${\tt {\it f}_{12}/{\it f}_{4.6})\gtrsim}$0.7-0.8 could be a
mixture of heavily absorbed AGN and objects at high redshifts
(\red$\gtrsim$1-1.5).

\begin{table}
  \begin{center}
    \caption{Comparison of our AGN wedge with other \wise selection techniques.}
    \label{tab5}
    \begin{tabular}{ccccc}
      \hline
      \hline
       MIR wedge & ${\rm N_{wedge}}$ & ${\rm N_{wedge+X}}$  \\
      (1) & (2) & (3)  \\
      \hline
      Mateos 3-band           & 2755   &  1062 (38.5\%)                      \\  
      Jarrett+11              & 3301   &  1102 (33.4\%)                      \\  
      Stern+12                & 3946  &  1254 (31.8\%)  \\
      \hline
    \end{tabular} \\
    Column 1: MIR AGN selection criteria; Column 2: number of
    catalogued \wise AGN candidates in the \buxs survey area. For the
    \citet{jarrett11} wedge we selected only MIR objects with
    significance of detection $\geq$5 in the three shorter wavelength
    bands of \wise as in our analysis. The \citet{stern12}
      selection only requires an MIR detection at 4.6\,$\mu$m,
      brighter than 160\,$\mu$Jy (see Sec.~\ref{comparison} for
      details). Column 3: number (fraction) of \wise sources in AGN
    wedge with an X-ray detection in the 2-10 keV band.
  \end{center}
\end{table}

\subsection[]{Comparison with other \wise selection techniques}
\label{comparison}
Within the \buxs survey area our three-band MIR colour selection
identifies 2755 AGN candidates, of which 1062 sources (38.5\%) are
detected in X-rays. For comparison, the X-ray detection fraction of
\wise objects that meet the \citet{jarrett11} colour-based AGN
selection (indicated in Fig.~\ref{fig2}) is 33.4\% (see Table \ref{tab5}). 
At red MIR colours their selection enters the sequence of low redshift
normal galaxies increasing the expected number of
contaminants. 

\citet{stern12} proposed an AGN selection using a [3.4]$-$[4.6] colour
cut ([3.4]$-$[4.6]$\geq$0.8 or {\tt log}(${\tt {\it f}_{4.6}/{\it
    f}_{3.4})\gtrsim 0.06}$) and a 4.6\mic flux threshold of
  160$\mu$Jy. Their argument was that the inclusion of the longer
wavelength \wise data would increase the reliability of the AGN
selection but at the cost of reducing the completeness.
Fig.~\ref{fig2} and Fig.~\ref{fig6} show that the [3.4]-[4.6] colour
cut proposed by \citet{stern12} also enters the locus of low redshift
normal galaxies at red {\tt log}(${\tt {\it f}_{12}/{\it f}_{4.6})}$
colours. This could reduce the reliability of their selection as
  suggested by the lower X-ray detection fraction of \wise sources in
  the \buxs area that meet their criteria (31.8\%; see Table
  \ref{tab5}). On the other hand, using the deep (60-160~ks) \chandra
  data available in the COSMOS field, \citet{stern12} find that 87\%
  of of their \wise AGN candidates are detected at X-ray energies,
  suggesting minimal contamination from normal galaxies.

The \xmm pointings used to build \buxs span a broad range of
  ecliptic latitudes and thus the depth of the \wise survey varies
  across the \buxs fields. The \citet{stern12} AGN selection was
  defined using the COSMOS field located at low ecliptic latitude and
  thus, the MIR data are close to the minimum depth of the \wise
  survey. At such shallow depths and using the \citet{stern12}
  160$\mu$Jy flux threshold at 4.6\nmic, many of the star-forming
  contaminants will be too faint to be detected, thereby increasing
  the reliability of the \citet{stern12} selection. Although \buxs
likely samples fainter objects than those best targeted at the shallow
depth of {\rm WISE}, we find that $\sim$98\% of the AGN have MIR
detections with \sn$\geq$5 at 3.4\mic and 4.6{$\mu$m}. This fraction
only decreases to $\sim$77\% if we require 12\mic detections with
\sn$\geq$5. In conclusion, over the range of MIR depths of the
  \wise survey in the \buxs fields, our proposed selection suffers less
  contamination from star-forming galaxies than provided by a simple
  [3.4]$-$[4.6] color cut, while only marginally reducing
  completeness.

We have investigated the impact of increasing the MIR significance of
detection of the AGN candidates. If we require 3.4\mic and 4.6\mic
detections with \sn$\geq$10, 44\% of our MIR AGN candidates with {\tt
  log}(${\tt {\it f}_{12}/{\it f}_{4.6}}$)$\geq$0.6 will be missed
(21\% of the objects with an X-ray detection), while this fraction is
only 0.7\% at {\tt log}(${\tt {\it f}_{12}/{\it f}_{4.6}}$)$\leq$0.6
(and zero for objects with an X-ray detection). Therefore, the effect
of increasing the threshold in \sn\, of the detections is that the
resulting MIR selection becomes increasingly similar to a hard X-ray
selection. By using \sn$\geq$5 \wise sources, we gain in the
identification of AGN candidates with the reddest MIR colours, while
we do not reduce the reliability. 

It is clear that our MIR colour selection is a good compromise between
completeness and the crucial high reliability required to obtain a
clean sample of powerful AGN at the different depths of the \wise
  survey.  Furthermore, going down to detections with \sn$\geq$5, we
reach a much higher efficiency of detection of the AGN population in
the reddest MIR colours, many of which could be heavily
obscured/extincted AGN.

\begin{figure}
  \centering
  \begin{tabular}{cc}
    \hspace{-0.7cm}\includegraphics[angle=90,width=0.49\textwidth]{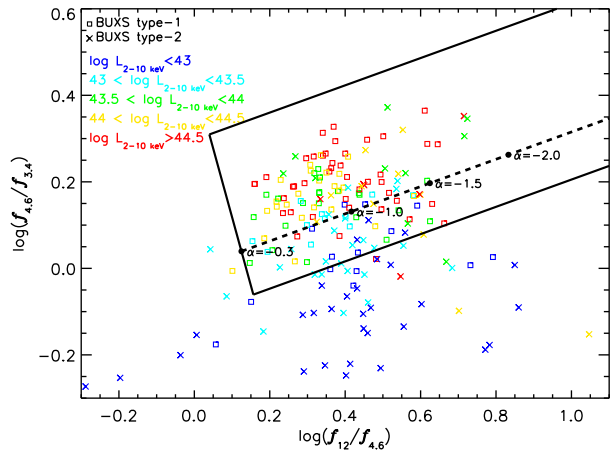}
  \end{tabular}
  \caption{MIR colours of \buxs AGN as a function of intrinsic
    2-10 keV luminosity. Solid and dashed lines illustrate the AGN
    selection wedge and power-law locus, respectively. Most \buxs 
type-2 AGN are objects with ${\tt {\it
        L}_{2-10\,keV}\lesssim 10^{44}\lum}$ ($\sim$79\% vs. $\sim$39\%
    for type-1 AGN; see Fig.~\ref{fig3}). At such luminosities many AGN have
    relatively blue colours at the shortest \wise wavelengths (i.e.,
    host-dominated) and lie outside the AGN wedge.}

  \label{fig4}
\end{figure}

\subsection[]{Trends in the wedge with the AGN luminosity}
In the previous sections we presented the \wise expected colours for
AGN/star-forming galaxies derived from a library of templates. Here we
investigate in more detail the colour trends in the three-band AGN
wedge for the \buxs type-1 and type-2 AGN.  Fig.~\ref{fig4} shows the
MIR colours of the AGN in \buxs as a function of their X-ray
luminosity. As expected, there is a strong dependence of {\tt
  log}(${\tt {\it f}_{4.6}/{\it f}_{3.4}}$) on the X-ray luminosity,
with less luminous sources having the bluest MIR colours. For less
powerful AGN the host galaxy can substantially contribute to the MIR
emission (e.g. \citealt{buchanan06}; \citealt{alonso08}). Host galaxy
dilution is expected to be more important in type-2 AGN, especially at
low luminosities, as type-2 AGN should show a higher degree of
extinction at the shortest wavelengths of {\rm WISE}. These objects,
with MIR colours consistent with normal galaxies, lie outside of the
AGN wedge and thus, they are missed from a pure MIR
selection. Fig.~\ref{fig2} and Fig.~\ref{fig4} both show that most AGN in
\buxs missed by the 3-band wedge are type-2 AGN. As noted in
Sec.~\ref{completeness}, in flux-limited X-ray surveys, such as
{\texttt{BUXS}}, type-2 AGN are overall intrinsically less luminous
than type-1 AGN (see Fig.~\ref{fig1}, bottom). Thus, due to the strong
dependence of the MIR selection completeness on the luminosity of the
objects (see Fig.~\ref{fig3}, top), our AGN wedge will preferentially
pick out \buxs type-1 AGN. However, at luminosities ${\tt {\it
    L}_{2-10\,keV}>10^{44}\lum}$, where the AGN is expected to
dominate the MIR emission (unless it is heavily absorbed), the
completeness of the selection of type-1 and type-2 AGN is comparable
within the uncertainties. At such luminosities both type-1 and type-2
AGN are preferentially located above the power-law locus (see also
Fig.~\ref{fig6}). This indicates that the observed 3.4\mic to 12\mic
MIR SEDs of powerful AGN deviate from a pure power-law.

We do not find a strong dependence of the {\tt log}(${\tt {\it
    f}_{12}/{\it f}_{4.6}}$) colour on the X-ray luminosity. However,
we find a large scatter in the distribution of {\tt log}(${\tt {\it
    f}_{12}/{\it f}_{4.6}}$) colours, especially for objects at low
luminosities, where the host galaxy significantly contributes to the
MIR emission. A broad range of {\tt log}(${\tt {\it f}_{12}/{\it
    f}_{4.6}}$) colours is expected for AGN with an important host
galaxy contribution as the very wide 12\mic filter of \wise is very
sensitive to both prominent PAH emission features and silicate
absorption (10{$\mu$m}) in star-forming galaxies over a broad range of
redshifts.

\section{SUMMARY}
We present a MIR power-law based selection of luminous AGN candidates
using the 3.4, 4.6, and 12 \mic bands of the \wise survey. We defined
an AGN wedge in the {\tt log}(${\tt {\it f}_{4.6}/{\it f}_{3.4}}$)
vs. {\tt log}(${\tt {\it f}_{12}/{\it f}_{4.6}}$) colour-colour
diagram using the Bright Ultra-Hard \xmm Survey
({\texttt{BUXS}}). This is one of the largest complete flux-limited
samples of bright (${\tt {\it f}_{4.5-10\,keV} > 6\,x\,10^{-14}
  \flux}$) ``ultra-hard'' (4.5-10 keV) X-ray selected AGN to
date. \buxs includes 258 objects detected over a total sky area of
44.43 deg$^2$: 251 (97.3\%) are spectroscopically identified and
classified, with 145 being type-1 AGN and 106 type-2 AGN. Our
technique is based on a MIR power-law selection and properly accounts
for the errors in the photometry and deviations of the MIR spectral
energy distributions from a pure power-law.  In flux-limited X-ray
surveys, such as {\texttt{BUXS}}, type-2 AGN are intrinsically less
luminous than type-1 AGN. Thus, due to the strong dependence of the
MIR selection completeness on the luminosity of the objects, a MIR AGN
wedge necessarily picks out \buxs type-1 AGN. However, at 2-10 keV
luminosities above ${\rm 10^{44}\lum}$ the completeness of our MIR
selection of type-1 and type-2 AGN is high and comparable for both
types within the uncertainties. Our selection is highly complete at
luminosities ${\tt {\it L}_{2-10\,keV}>10^{44}\lum}$ where our MIR
wedge recovers $\sim$97\% and $\sim$77\% of the \buxs type-1 and
type-2 AGN, respectively. We identify 2755 AGN candidates in the 44.43
deg$^2$ \buxs survey area of which 38.5\% have detection in X-rays. In
the \buxs area where the X-ray observations have exposures $>$40\,ks,
the X-ray detection fraction rises to 49.8\%. This is reasonable, as
long X-ray exposures are required to detect intrinsically less
luminous and/or heavily obscured AGN. A substantial fraction of the
MIR AGN candidates remain undetected at 2-10\,keV energies with the
typical exposures in the {\tt 2XMM} catalogue. These objects are the
best candidates to account for the most heavily obscured/absorbed
luminous AGN missed by hard X-ray surveys. Assuming that a 2-10 keV
X-ray detection is a good tracer of AGN activity we demonstrate that
our \wise selection shows one of the highest reliability amongst those
in the literature. This is crucial to obtain a clean MIR selection
of powerful AGN. Furthermore, going down to a \sn$\geq$5 limit in the
\wise flux densities, we substantially increase the efficiency of
detection of AGN with the reddest MIR colours. We also investigate a
\wise four-band AGN selection. We show, however, that by including the
22\mic \wise band neither the completeness nor the reliability of the
selection improves. This is likely due to both the significantly
shallower depth at 22\mic compared with the first three bands of \wise
and star-formation contributing to the 22\mic emission at the \wise
22\mic sensitivity. \\

\section*{Acknowledgments}
This work is based on observations obtained with
{\textit{XMM-Newton}}, an ESA science mission with instruments and
contributions directly funded by ESA Member States and NASA. Based on
data from the Wide-field Infrared Survey Explorer, which is a joint
project of the University of California, Los Angeles, and the Jet
Propulsion Laboratory/California Institute of Technology, funded by
the National Aeronautics and Space Administration.  Funding for the
SDSS and SDSS-II has been provided by the Alfred P. Sloan Foundation,
the Participating Institutions, the National Science Foundation, the
U.S. Department of Energy, the National Aeronautics and Space
Administration, the Japanese Monbukagakusho, the Max Planck Society,
and the Higher Education Funding Council for England. The SDSS Web
Site is http://www.sdss.org/.  Based on observations collected at the
European Organisation for Astronomical Research in the Southern
Hemisphere, Chile, programme IDs 084.A-0828, 086.A-0612,
087.A-0447. Based on observations made with the William Herschel
Telescope -operated by the Isaac Newton Group-, the Telescopio
Nazionale Galileo -operated by the Centro Galileo Galilei and the Gran
Telescopio de Canarias installed in the Spanish Observatorio del Roque
de los Muchachos of the Instituto de Astrofísica de Canarias, in the
island of La Palma. SM, FJC and XB acknowledge financial support by
the Spanish Ministry of Economy and Competitiveness through grant
AYA2010-21490-C02-01. A.A.-H. acknowledges support from the
Universidad de Cantabria through the Augusto G. Linares program.
A.B. acknowledges a Royal Society Wolfson Research Merit Award.
J.L.D. acknowledges support from the LANL Director's Fellowship.
P.S. acknowledges financial support from ASI (grant No. I/009/10/0).
The authors wish to thank the anonymous referee for constructive
comments.

\small
\bibliographystyle{mn2e}

\bsp

\label{lastpage}

\end{document}